# Ultra-slow light pulses in a nonlinear metamaterial

**Giuseppe D'Aguanno†, Nadia Mattiucci and Mark J. Bloemer**
*Charles M. Bowden Research Center, RDECOM, Building 7804, Redstone Arsenal, AL 35898-5000, USA.*

**Abstract**

We find the analytical expression for the threshold intensity necessary to launch ultra-slow light pulses in a metamaterial with simultaneous cubic electric and magnetic nonlinearity. The role played respectively by the permittivity, the permeability, the electric cubic nonlinearity, the magnetic cubic nonlinearity and the pulse duration is clearly identified and discussed.

† Corresponding Author: giuseppe.daguanno@us.army.mil *or* giuseppe.daguanno@gmail.com; tel. 001-256-8429815; fax 001-256-8422507

## 1. Introduction

Temporal solitons, i.e. guided light pulses which propagate without dispersion due to the balancing between the group-velocity dispersion (GVD) and the self-phase modulation induced by a Kerr nonlinearity, play a fundamental role in optical communications systems [1,2]. In the past few years, metamaterials, i.e. artificial composites assembled in such a way that they show both an effective electric and magnetic response although the constituent materials are non-magnetic, have been the subject of intense theoretical and experimental investigations due to their vast range potential applications from super-resolution [3] to cloaking [4]. The study of optical solitons in metamaterials is a new and exciting field of research which has already produced some important theoretical results: we cite for example the possibility to excite bright and dark gap solitons in a metamaterial cavity [5], a new nonlinear Schrödinger equations (NLSE) for metamaterials with only cubic electric nonlinearity [6], a generalized system of two, coupled, spatio-temporal NLSEs for metamaterials with simultaneous cubic electric and magnetic nonlinearity [7].

The aim of this paper is to arrive to an analytical expression for the threshold intensity needed to launch ultra-slow optical soliton in a metamaterial and to clearly put into evidence the role played respectively by the permittivity, the permeability, the electric cubic nonlinearity, the magnetic cubic nonlinearity and the pulse duration. “Slow-light” has recently received a great deal of attention in telecommunications for its numerous applications ranging from all-optical storage to all-optical switching [8]. Before going into the mathematical analysis of the problem we would like to discuss briefly the linear properties of a metamaterial which we describe by a Drude model [3] for both permittivity and permeability: $\varepsilon(\tilde{\omega}) \approx 1 - 1/\tilde{\omega}^2$, $\mu(\tilde{\omega}) \approx 1 - (\omega_{mp}/\omega_{ep})^2/\tilde{\omega}^2$, where $\tilde{\omega} = \omega/\omega_{ep}$ is the normalized frequency, $\omega_{ep}$ and $\omega_{mp}$ are respectively the electric plasma frequency and the magnetic plasma frequency of the metamaterial and $\omega_{ep}=2\pi c/\lambda_{ep}$ where $\lambda_{ep}$ is the electric plasma wavelength. In Fig.1(a) we show the refractive index $n = \pm\sqrt{\varepsilon\mu}$ and group velocity (GV) $V_g$ for $\omega_{mp}/\omega_{ep} = 0.8$. The negative determination of the square root must be taken when $\varepsilon$ and $\mu$ are simultaneously negative [9]. The material is characterized by an opaque region (no propagative modes allowed) between $\omega_{mp}<\omega<\omega_{ep}$ and two transparent regions (propagative modes allowed) respectively for $\omega<\omega_{mp}$ and $\omega>\omega_{ep}$, note also that $V_g$ and $n$ reach values close to zero near the edges of the propagative regions. The two band edges are defined respectively by the conditions $\omega=\omega_{ep}$ and $\omega=\omega_{mp}$ . It is indeed near the band edges of the propagative regions that ultra-slow GV temporal solitons make their appearance. In Fig.1(b) we show the so-called GVD parameter $\beta_2 \equiv \left(\frac{d}{d\omega}\left(\frac{1}{V_g}\right)\right)$. The zones at the edges of the propagative region where low GV is achieved are also characterized by a strong dispersion that, as we will see later, substantially increases the threshold intensity necessary to launch the fundamental soliton. We also note that in the case of an impedance matched metamaterial, i.e. $\omega_{mp} = \omega_{ep}$, the opaque region disappears and, as a consequence, the GV remains substantially close to $c$ over all the spectrum.

The paper is organized as follows: in Section 2 we derive the basic equations and arrive to a system of two NLSEs that couple together the electric and magnetic field, in Section 3 we discuss the analytical solutions and in particular we focus on the ultra-slow soliton solutions near the band edges of the metamaterial, in Section 4 we go to the conclusions.

## 2. Basic Equations

Now, let us start our analysis with the Maxwell equations in (1+1)dimension for linearly polarized fields: $\frac{\partial E}{\partial z} = -\frac{\partial B}{\partial t}$ , $-\frac{\partial H}{\partial z} = \frac{\partial D}{\partial t}$ where E, B, H and D are respectively the electric field (E), the magnetic induction (B), the magnetic field (H) and the electric displacement or electric induction (D). In what follows we use non dimensional units, i.e. $\varepsilon_0 = \mu_0 = c = 1$, where $\varepsilon_0$ and $\mu_0$ are respectively the permittivity and permeability in vacuo and $c$ is the speed of light in vacuo. We write the electric and magnetic field as the product of an envelope function multiplied a harmonic oscillation at frequency $\omega_0$: $E(z,t) = (1/2)[\tilde{E}(z,t)e^{-i\omega_o t} + c.c]$ and $H(z,t) = (1/2)[\tilde{H}(z,t)e^{-i\omega_o t} + c.c]$. In the hypothesis that the envelope of the fields is slowly varying in time with respect to the oscillation associated with the carrier frequency $\omega_0$, the time derivatives respectively over B and D can be approximated as follows:

$$\frac{\partial D}{\partial t} \cong \frac{1}{2} e^{-i\omega_0 t}\left[ -i\omega_0 \varepsilon \tilde{E} + \left(\frac{\partial(\varepsilon\omega)}{\partial\omega}\right)_{\omega=\omega_0} \frac{\partial\tilde{E}}{\partial t} + \frac{i}{2}\left(\frac{\partial^2(\varepsilon\omega)}{\partial\omega^2}\right)_{\omega=\omega_0} \frac{\partial^2\tilde{E}}{\partial t^2} - i\omega_0 \varepsilon \tilde{\chi}_\varepsilon^{(3)} \left|\tilde{E}\right|^2 \tilde{E} \right] + c.c.$$

and

$$\frac{\partial B}{\partial t} \cong \frac{1}{2} e^{-i\omega_0 t}\left[ -i\omega_0 \mu \tilde{H} + \left(\frac{\partial(\mu\omega)}{\partial\omega}\right)_{\omega=\omega_0} \frac{\partial\tilde{H}}{\partial t} + \frac{i}{2}\left(\frac{\partial^2(\mu\omega)}{\partial\omega^2}\right)_{\omega=\omega_0} \frac{\partial^2\tilde{H}}{\partial t^2} - i\omega_0 \mu \tilde{\chi}_\mu^{(3)} \left|\tilde{H}\right|^2 \tilde{H} \right] + c.c.$$

where $\tilde{E}$ e $\tilde{H}$ are the complex envelopes respectively for the electric and magnetic fields, ε and μ are respectively the relative permittivity and the permeability of the material, permittivity and permeability which we suppose real quantities, $\tilde{\chi}_\varepsilon^{(3)} = (3/4)\chi_\varepsilon^{(3)}$ and $\tilde{\chi}_\mu^{(3)} = (3/4)\chi_\mu^{(3)}$ are respectively the cubic electric and magnetic nonlinearity which we suppose for simplicity non-dispersive. In the derivation of the approximated expressions for $\partial D/\partial t$ and $\partial B/\partial t$ we have neglected: a) the third order time derivative over the field envelopes; b) the first order time derivative over the nonlinear terms, i.e. the self-steepening terms like $\partial(|\tilde{E}|^2\tilde{E})/\partial t$ and $\partial(|\tilde{H}|^2\tilde{H})/\partial t$; c) the nonlinear terms whose temporal oscillations are different from the oscillation at the carrier frequency (rotating wave approximation), i.e. we retain only the self-phase modulation terms. Now, if we put into evidence the fundamental spatial oscillation by writing $\tilde{E}=\hat{E}e^{ikz}$ and $\tilde{H}=\hat{H}e^{ikz}$, where $k=n\omega_0$ is the fundamental wave-vector, and substitute the above expressions into Maxwell equations, we arrive to the following system of coupled equations:

$$\frac{\partial\hat{E}}{\partial z}+ik\hat{E}=i\omega_0(\mu+\tilde{\chi}_\mu^{(3)}|\hat{H}|^2)\hat{H}-\beta_{1,\mu}\frac{\partial\hat{H}}{\partial t}-\frac{i}{2}\beta_{2,\mu}\frac{\partial^2\hat{H}}{\partial t^2}\quad, \tag{1.1}$$

$$\frac{\partial\hat{H}}{\partial z}+ik\hat{H}=i\omega_0(\varepsilon+\tilde{\chi}_\varepsilon^{(3)}|\hat{E}|^2)\hat{E}-\beta_{1,\varepsilon}\frac{\partial\hat{E}}{\partial t}-\frac{i}{2}\beta_{2,\varepsilon}\frac{\partial^2\hat{E}}{\partial t^2}\quad, \tag{1.2}$$

where $\beta_{m,\varepsilon}\equiv\left(\frac{d^m(\omega\varepsilon)}{d\omega^m}\right)_{\omega=\omega_0}$ e $\beta_{m,\mu}\equiv\left(\frac{d^m(\omega\mu)}{d\omega^m}\right)_{\omega=\omega_0}$ (*m=1,2,…*). Eqs(1) have an evident symmetry, in fact Eq.(1.2) can be obtained from Eq.(1.1) and vice versa with the formal substitutions: ε→μ , μ→ε, $\hat{E}\rightarrow\hat{H}$, $\hat{H}\rightarrow\hat{E}$. We now concentrate on the manipulation of only one of the two equations. Let us concentrate on Eq.(1.1). By deriving Eq.(1.1) with respect to *z*, using Eq.(1.2) and neglecting the third order spatio-temporal derivatives over the envelopes and the first order derivatives over the nonlinear terms we arrive to the following equation:

$$\begin{aligned}&\frac{1}{2n\omega_0}\frac{\partial^2\hat{E}}{\partial z^2}+i\frac{\partial\hat{E}}{\partial z}+\frac{i(\varepsilon\beta_{1,\mu}+\mu\beta_{1,\varepsilon})}{2n}\frac{\partial\hat{E}}{\partial t}-\left(\frac{\beta_{1,\varepsilon}\beta_{1,\mu}}{2n\omega_0}+\frac{\beta_{2,\varepsilon}\mu}{4n}+\frac{\beta_{2,\varepsilon}n}{4\mu}\right)\frac{\partial^2\hat{E}}{\partial t^2}+\\&+\frac{\omega_0}{2}\left(\frac{\mu}{n}\tilde{\chi}_\varepsilon^{(3)}|\hat{E}|^2\hat{E}+\tilde{\chi}_\mu^{(3)}|\hat{H}|^2\hat{H}\right)=0\end{aligned} \tag{2}$$

Note that by neglecting the third order spatio-temporal derivatives we are implicitly assuming that the electromagnetic pulse is slowly varying both in time and space. While in standard positive index materials (n>1, μ=1) the condition that the pulse is slowly varying in time generally implies that the pulse is also slowly varying in space, in our case the situation is not so simple because of the high dispersion present in the propagative regions near the electric or the magnetic plasma frequency of the metamaterial (see Fig.1), i.e. where slow GV solitons are present. In general the condition of slowly varying envelope in space is broken anytime the pulse extension in the material $\Delta z=V_g T_0$ ($T_0$ is the temporal pulse duration) becomes comparable with the wavelength of the pulse $\lambda=2\pi c/(n\omega_0)$, in other words Eq.(2) retains its validity when $T_0>>2\pi c/(V_g n\omega_0)$. It is clear that this condition put a lower limit on how short in time a low GV soliton can be. We will come later to a quantitative analysis of this condition. For the time being, let us continue with the manipulation of Eq.(2). By defining $\beta_m \equiv \left(\frac{d^m(n\omega)}{d\omega^m}\right)_{\omega=\omega_0}$ (*m=1,2…*), by using the typical coordinate transformation: $\xi = z,\ \tau = t-\beta_1 z$, and finally by neglecting the terms which are of the same order as the third order spatio-temporal derivatives over the envelopes and the first order derivatives over the nonlinear terms, Eq.(2) can be put in the following form:

$$i\frac{\partial \hat{E}}{\partial \xi} - \frac{\beta_2}{2}\frac{\partial^2 \hat{E}}{\partial \tau^2} + \frac{\omega_0}{2}\left(\frac{\mu}{n}\tilde{\chi}_\varepsilon^{(3)}\left|\hat{E}\right|^2\hat{E} + \tilde{\chi}_\mu^{(3)}\left|\hat{H}\right|^2\hat{H}\right) = 0 \quad . \tag{3.1}$$

In arriving to Eq.(3.1) we have also used the following equalities: $\beta_1 = (\varepsilon\beta_{1,\mu} + \mu\beta_{1,\varepsilon})/2n$ and $\beta_2 = -\frac{1}{n\omega_0}\left(\frac{1}{V_g^2} - \beta_{1,\varepsilon}\beta_{1,\mu} - \frac{\beta_{2,\varepsilon}\omega_0\mu}{2} - \frac{\beta_{2,\mu}\omega_0\varepsilon}{2}\right)$ which can be demonstrated. We note that both $\beta_1$ and $\beta_2$ are invariant under the transformations: ε→μ,

μ→ε. Physically this means that the electric and magnetic field obviously must have the same GV and the same GVD. Eq.(3.1) is the first of the two equations we were looking for. It becomes the standard NLSE [1,2] for $\tilde{\chi}_{\mu}^{(3)} = 0$. Now, using the same procedure outlined above, but starting by deriving Eq.(1.2) with respect to *z*, we can arrive to the second equation:

$$i\frac{\partial \hat{H}}{\partial \xi} - \frac{\beta_2}{2}\frac{\partial^2 \hat{H}}{\partial \tau^2} + \frac{\omega_0}{2}\left(\frac{\varepsilon}{n}\tilde{\chi}_{\mu}^{(3)}\left|\hat{H}\right|^2\hat{H} + \tilde{\chi}_{\varepsilon}^{(3)}\left|\hat{E}\right|^2\hat{E}\right) = 0 \quad . \tag{3.2}$$

Eq.(3.1) and (3.2) form a system of two coupled NLSEs that preserve the same symmetry of Eqs(1), in fact we can formally obtain Eq.(3.2) from Eq.(3.1) and vice versa with the formal substitutions: ε→μ , μ→ε, $\hat{E} \to \hat{H}$, $\hat{H} \to \hat{E}$. Before going to analyze the analytical solutions of Eqs.(3), we make some cosmetic manipulations by introducing the following new variables and parameters: $\tilde{z} = \xi / L_D$, $L_D = T_0^2 / |\beta_2|$, $\tilde{t} = \tau / T_0$, $u_1 = \sqrt{L_D \omega_0 / 2}\hat{E}$, $u_2 = \sqrt{L_D \omega_0 / 2}\hat{H}$, $Z = \sqrt{\mu / \varepsilon}$, where $T_0$ is the pulse duration, $L_D$ is the so-called second-order dispersion length [1,2] and *Z* is the *impedance* of the medium which obviously is always positive. In the new variables, Eqs.(3) become:

$$i\frac{\partial u_1}{\partial \tilde{z}} - \frac{\mathrm{sgn}(\beta_2)}{2}\frac{\partial^2 u_1}{\partial \tilde{t}^2} + \left(Z\tilde{\chi}_{\varepsilon}^{(3)}|u_1|^2 u_1 + \tilde{\chi}_{\mu}^{(3)}|u_2|^2 u_2\right) = 0 \quad , \tag{4.1}$$

$$i\frac{\partial u_2}{\partial \tilde{z}} - \frac{\mathrm{sgn}(\beta_2)}{2}\frac{\partial^2 u_2}{\partial \tilde{t}^2} + \left(\frac{1}{Z}\tilde{\chi}_{\mu}^{(3)}|u_2|^2 u_2 + \tilde{\chi}_{\varepsilon}^{(3)}|u_1|^2 u_1\right) = 0 \quad , \tag{4.2}$$

where $\mathrm{sgn}(\beta_2)$ stands for the sign of the GVD parameter.

## 3. Ultra-slow solitons

The fundamental soliton solutions of Eqs(4) can be easily found by noting that *this system of two coupled NLSEs can be decoupled into a single NLSE by using the following transformations:* $u_1 = Zu$, $u_2 = u$. The transformations used to decouple the system of Eq.(4)

have an obvious physical meaning: the electric and the magnetic field must be proportional each other through the impedance $Z$ of the medium, as one may expect. The single NLSE in the $u$ variable takes the following form:

$$i\frac{\partial u}{\partial \tilde{z}} - \frac{\mathrm{sgn}(\beta_2)}{2}\frac{\partial^2 u}{\partial \tilde{t}^2} + \Gamma |u|^2 u = 0 \quad , \qquad (5)$$

where $\Gamma = \left(Z^3 \tilde{\chi}_\varepsilon^{(3)} + (1/Z)\tilde{\chi}_\mu^{(3)}\right)$. The fundamental soliton solutions of Eq.(5) are well known [1,2].

Bright soliton: $u(\tilde{z},\tilde{t}) = (\mathrm{sec}h(\tilde{t})/\sqrt{|\Gamma|})\exp(\pm i\tilde{z}/2)$ for $\mathrm{sgn}(\beta_2) = -1$ and $\Gamma > 0$ or $\mathrm{sgn}(\beta_2) = 1$ and $\Gamma < 0$. Dark soliton: $u(\tilde{z},\tilde{t}) = (\tanh(\tilde{t})/\sqrt{|\Gamma|})\exp(\pm i\tilde{z})$ for $\mathrm{sgn}(\beta_2) = 1$ and $\Gamma > 0$ or $\mathrm{sgn}(\beta_2) = -1$ and $\Gamma < 0$. The threshold intensity to launch the fundamental solitons can be calculated from the maximum of the Poynting vector, $S_{thr} = (1/2)\mathrm{Re}\left[\tilde{E}\tilde{H}^*\right]_{Max}$ which in our case gives:

$$S_{thr} = \frac{Z}{L_D \omega_0 |\Gamma|} = \frac{|\beta_2|\varepsilon\mu}{T_0^2 \omega_0 \left|\mu^2 \tilde{\chi}_\varepsilon^{(3)} + \varepsilon^2 \tilde{\chi}_\mu^{(3)}\right|} \quad . \qquad (6)$$

Eq.(6) is the analytical expression for the threshold intensity necessary to launch the fundamental solitons in a generic metamaterial having both electric and magnetic cubic nonlinearity. In order to analyze quantitatively the situation for simplicity we assume that the electric and magnetic nonlinearity are of the same order of magnitude. By tuning the carrier frequency of the pulse near the electric plasma frequency, i.e. $\omega_0 \approx \omega_{ep}$ and $\varepsilon$~0, we obtain from Eq.(6) the following expression: $S_{thr} \approx |\beta_2|(\varepsilon/\mu)/\left(T_0^2 \omega_0 |\tilde{\chi}_\varepsilon^{(3)}|\right)$, which means that the ultra-slow GV soliton is of "electrical" nature, namely it is the electric nonlinearity which plays the dominant role for its formation while the magnetic nonlinearity is quenched. Vice versa, near the magnetic plasma frequency, i.e. $\omega_0 \approx \omega_{mp}$ and $\mu$~0, we obtain from Eq.(6)

$S_{thr} \approx |\beta_2|(\mu/\varepsilon)/\left(T_0^2\omega_0|\tilde{\chi}_\mu^{(3)}|\right)$, i.e. this time it is the magnetic nonlinearity to play a dominant role in the formation of the ultra-slow GV soliton . It is worthwhile to remark that in the above discussion we are operating *in regions near the band edges, but not exactly at the band edges*; in other words, the carrier frequency of the pulse is always slightly detuned with respect to the electric or magnetic plasma frequency of the metamaterial. This means that the refractive index and the permittivity or permeability are very small but not quite zero, avoiding therefore any intrinsic singularity in Eq.(2) or in any other equation that derives from Eq.(2). Finally in the case of an impedance matched metamaterial, i.e. $\omega_{mp} = \omega_{ep}$, we obtain $S_{thr} = |\beta_2|/\left(T_0^2\omega_0|\tilde{\chi}_\varepsilon^{(3)} + \tilde{\chi}_\mu^{(3)}|\right)$, which means that the electric and magnetic nonlinearity play an equal important role in the soliton formation, although in this case no ultra-slow GV soliton is excited. In order to have an idea of how much intensity is necessary to launch these ultra-slow GV light pulses we draw a comparison with the intensity necessary to launch a soliton in a standard fiber. For simplicity we fix the electric plasma wavelength of the metamaterial at $\lambda_{ep}=1\mu m$ ($1/\omega_{ep}\cong 0.55fs$). In a standard single mode fiber the dispersion $D=-(2\pi c/\lambda^2)\beta_2$ is approximately *D~15ps/(Km nm)* at ~1.55μm [1,2]. The GVD parameter of the standard fiber at 1.55μm expressed in units of $(c\omega_{ep})^{-1}$ is approximately $\beta_2^{fiber} \approx -10^{-2}(c\omega_{ep})^{-1}$ and threshold intensity is $S_{thr}^{fiber} \approx 3\cdot 10^{-2}/\left(T_0^2\omega_{ep}^2 c|\tilde{\chi}_{fiber}^{(3)}|\right)$ where we take $n_{fiber}\sim 1.4$. In Fig.2 we plot $(S_{thr}|\tilde{\chi}_{fiber}^{(3)}|)/(S_{thr}^{fiber}|\tilde{\chi}_\varepsilon^{(3)}|)$ and $V_g$ near the electric plasma frequency of the metamaterial. In this case $\beta_2$ is negative (see Fig.1(b)) and bright solitons can be launched when $\tilde{\chi}_\varepsilon^{(3)} > 0$. For $\omega_0=1.00002\omega_{ep}$ (i.e. a detuning from the high frequency band edge $\delta\omega = \omega_0 - \omega_{ep} = 2*10^{-5}\omega_{ep}$) the GV will be *~c/100* and $(S_{thr}|\tilde{\chi}_{fiber}^{(3)}|)/(S_{thr}^{fiber}|\tilde{\chi}_\varepsilon^{(3)}|) \sim 10^4$. Of course, the actual value of the intensity necessary to launch the soliton will depend on the value of the cubic nonlinearity in the metamaterial. It has been noted [10] that microscopic fields can be dramatically enhanced in a composite structure like a metamaterial giving therefore an enhancement of nonlinear phenomena. If we suppose: $\tilde{\chi}^{(3)} \sim (10^3 \div 10^4)\tilde{\chi}^{(3)}_{fiber}$, the intensity necessary to launch a *c/100* GV soliton will be comparable with the intensity necessary to launch a soliton in a standard fiber. The intensity grows exponentially if slower pulses are to be launched. As we have already mentioned, there is a limitation to how short a pulse can be. In the case of our example we have at $\omega_0=1.00002\omega_{ep}$ the following values: $n\sim 4*10^{-3}$ and $Vg\sim 10^{-2}c$, therefore the temporal duration of the pulse must be $T_0>>2\pi c/(V_g n\omega_0)\sim 100ps$, which means at least ~1*ns* pulses. There is one last issue that deserves to be discussed in some detail and this is the effect of higher order dispersion terms. As we have already remarked in the manuscript, we are operating near the band edges of the structure, i.e. in regions of extremely high dispersion, as it is evident, for example, from Fig.1(b) where the GVD parameter $\beta_2$ is represented. On the other hand, in order to arrive to analytical solutions, we have neglected the third and higher order dispersion terms. Now, in conventional fibers, third and higher order dispersion terms are generally small and can be treated by perturbation approaches for a pulse duration $T_0>>1ps$ [1,2]; in our case, given the remarkably high dispersion present near the band edges, we may expect that much longer pulses than those of a conventional fiber are needed in order to reduce the effects of the third and high order dispersion. We would like to remind the reader that we have already set up a lower limit to the pulse duration, i.e. $T_0>>2\pi c/(V_g n\omega_0)$, based on the requirement that the pulse must be slowly varying in space; in our case, for a soliton with a group velocity $10^{-2}c$, this condition calls for pulses that are at least *1ns* in duration. Let us here concentrate on the third order dispersion in particular. In order to consider the third order dispersion as a small perturbation we need to impose the condition that $L_D^{(3)} >> L_D$ where $L_D^{(3)} = T_0^3 / |\beta_3|$ is the so-called third-order dispersion length [1,2, 11] and $L_D = T_0^2 / |\beta_2|$ the second-order dispersion length. This means that the pulse duration must satisfy the following condition: $T_0 >> |\beta_3| / |\beta_2|$. In Fig.3 we plot $|\beta_3| / |\beta_2|$ for the metamaterial described in Fig.1 when the carrier frequency of the pulse is tuned near the high frequency band edge, i.e . $\omega_0 \sim \omega_{ep}$ . In particular we plot $|\beta_3| / |\beta_2|$ as function of the detuning $\delta\omega = \omega_0 - \omega_{ep}$ . In the case of a detuning $\delta\omega / \omega_{ep} \approx 2 * 10^{-5}$ the GV of the soliton is c/100 and $T_0 >> |\beta_3| / |\beta_2| \approx 7.4 * 10^4 / \omega_{ep} \sim 40 ps$. If we now recall that for the same tuning condition the pulse duration necessary to maintain the validity of the slowly varying envelope approximation is $T_0 >> 2\pi c/(V_g n \omega_0) \sim 100ps$, it should appear evident that *ns*-pulses not only ensure the validity of the slowly varying envelope approximation in space, but also allow us to treat the third order dispersion as a small perturbation. The generalization of Eq.(5) which includes the third order dispersion can be written as:

$$i \frac{\partial u}{\partial \tilde{z}} - \frac{\mathrm{sgn}(\beta_2)}{2} \frac{\partial^2 u}{\partial \tilde{t}^2} + \Gamma |u|^2 u = i \delta_3 \frac{\partial^3 u}{\partial \tilde{t}^3} \quad , \tag{7}$$

where $\delta_3$ is the third order dispersion parameter: $\delta_3 = \beta_3 / (6 T_0 |\beta_2|)$ and $|\delta_3| << 1$ in our case. Here we consider for simplicity the case that the pulse is tuned near the high frequency band edge and the electric nonlinearity is positive. It can be demonstrated by using a perturbation approach [2,13] that the fundamental bright soliton solution modified by the third order dispersion can be written in the following form:

$$u(\tilde{z}, \tilde{t}) = (\sec h(\tilde{t} - \delta_3 \tilde{z}) / \sqrt{\Gamma}) \exp(i\tilde{z} / 2) \quad . \tag{8}$$

In other words, what the third order dispersion does is to influence the actual GV of the bright soliton, GV which takes the following form:

$$V_g^{(1)} = \frac{V_g}{1 + \frac{V_g \beta_3}{6T_0^2}} \quad , \tag{9}$$

where clearly $V_g=1/\beta_1$ is the usual group velocity. In Fig.4 we calculate the correction brought to the usual group velocity by the third order dispersion for a *1ns*-pulse tuned near the high frequency band edge. The figure shows that the GV is corrected for less than one part over one thousand and therefore for all intents and purposes this correction can be considered negligible.

Similar results to those exposed above can be expected if $\omega_0$ is tuned near $\omega_{mp}$, except that in this latter case bright solitons can be launched when $\tilde{\chi}_\mu^{(3)} < 0$.Here we have explicitly analyzed the case: $\omega_{mp}<\omega_{ep}$. In the opposite situation: $\omega_{mp}>\omega_{ep}$, we would have bright solitons at $\omega_0 \approx \omega_{mp}$ for $\tilde{\chi}_\mu^{(3)} > 0$ and at $\omega_0 \approx \omega_{ep}$ for $\tilde{\chi}_\varepsilon^{(3)} < 0$. A final note regarding the absorption of the metamaterial. In this work we have neglected the absorption of the metamaterial. In currently available metamaterials in the near infrared [12] the absorption is still so high that it is premature to think about any practical soliton application. Nevertheless, in principle, there is nothing that prevents the possible availability of low-loss metamaterials in the near future.

**4. Conclusions**

In conclusion, we have performed an analytical study on the possibility to excite ultra-slow solitons near the band-edges of a metamaterial. We have investigated the role played respectively by the permittivity, the permeability, the electric cubic nonlinearity, the magnetic cubic nonlinearity and the pulse duration. We hope that our results may stimulate further research aiming at the study of this new class of materials in applications which involve “slow light”, such as all-optical buffering and switching for example.

**Acknowledgments**

G.D. and N.M. thank the National Research Council for financial support and Claudio Conti for helpful discussions.

**Figure Captions**

**Fig.1:** (a) $n$ (dashed line) and $V_g$ in units of $c$ (continuous line) vs. $\omega/\omega_{ep}$ for a metamaterial described by a Drude model with $\omega_{mp}/\omega_{ep}$ =0.8 as outlined in the main text. (b) GVD parameter $\beta_2$ vs. $\omega/\omega_{ep}$.

**Fig.2**: $(S_{thr}|\tilde{\chi}^{(3)}_{fiber}|)/(S^{fiber}_{thr}|\tilde{\chi}^{(3)}_{\varepsilon}|)$ (left y-axis) and $V_g$ in units of $c$ (right y-axis) vs. $\delta\omega/\omega_{ep}$ where $\delta\omega = \omega_0 - \omega_{ep}$ is the detuning from the high frequency band edge.

**Fig.3:** $|\beta_3|/|\beta_2|$ (units of $1/\omega_{ep}$) vs. $\delta\omega/\omega_{ep}$.

**Fig.4:** $|V_g^{(1)} - V_g|/V_g$ vs. $\delta\omega/\omega_{ep}$.

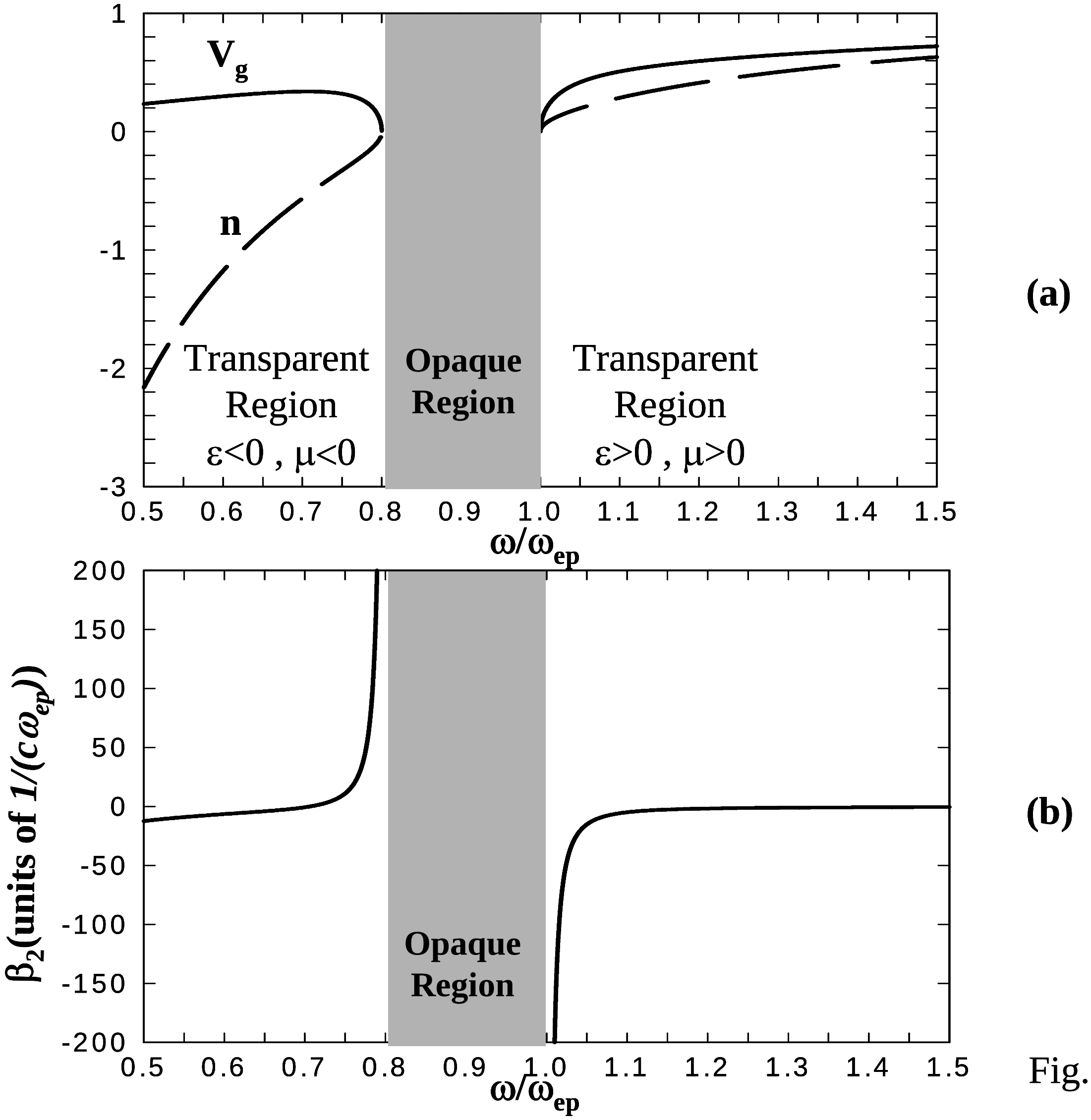

Fig.1

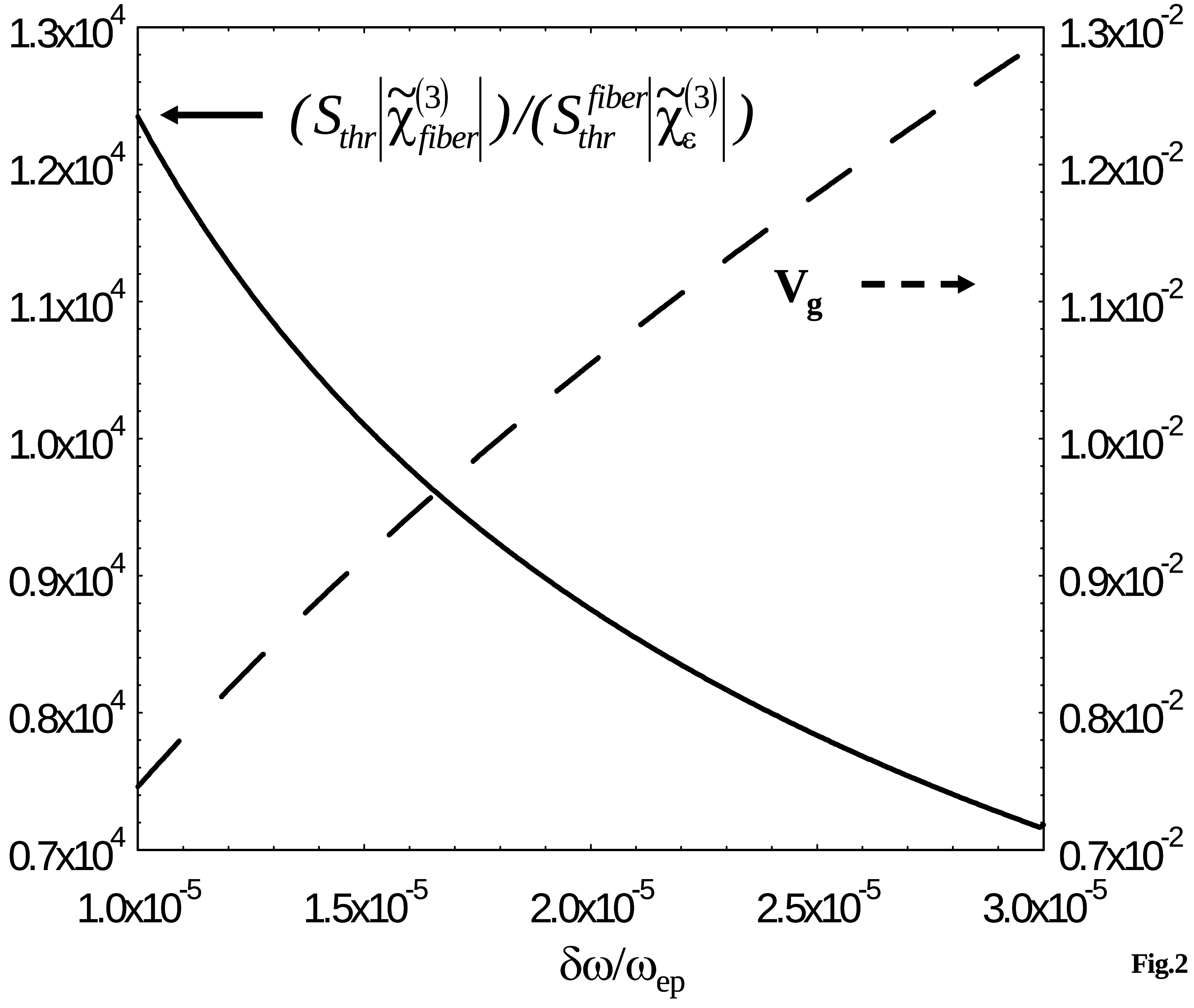

1.3x10^4
1.2x10^4
1.1x10^4
1.0x10^4
0.9x10^4
0.8x10^4
0.7x10^4
1.3x10^-2
1.2x10^-2
1.1x10^-2
1.0x10^-2
0.9x10^-2
0.8x10^-2
0.7x10^-2
$(S_{thr}|\tilde{\chi}^{(3)}_{fiber}|)/(S^{fiber}_{thr}|\tilde{\chi}^{(3)}_{\varepsilon}|)$
$V_g$
1.0x10^-5
1.5x10^-5
2.0x10^-5
2.5x10^-5
3.0x10^-5
$\delta\omega/\omega_{ep}$

**Fig.2**

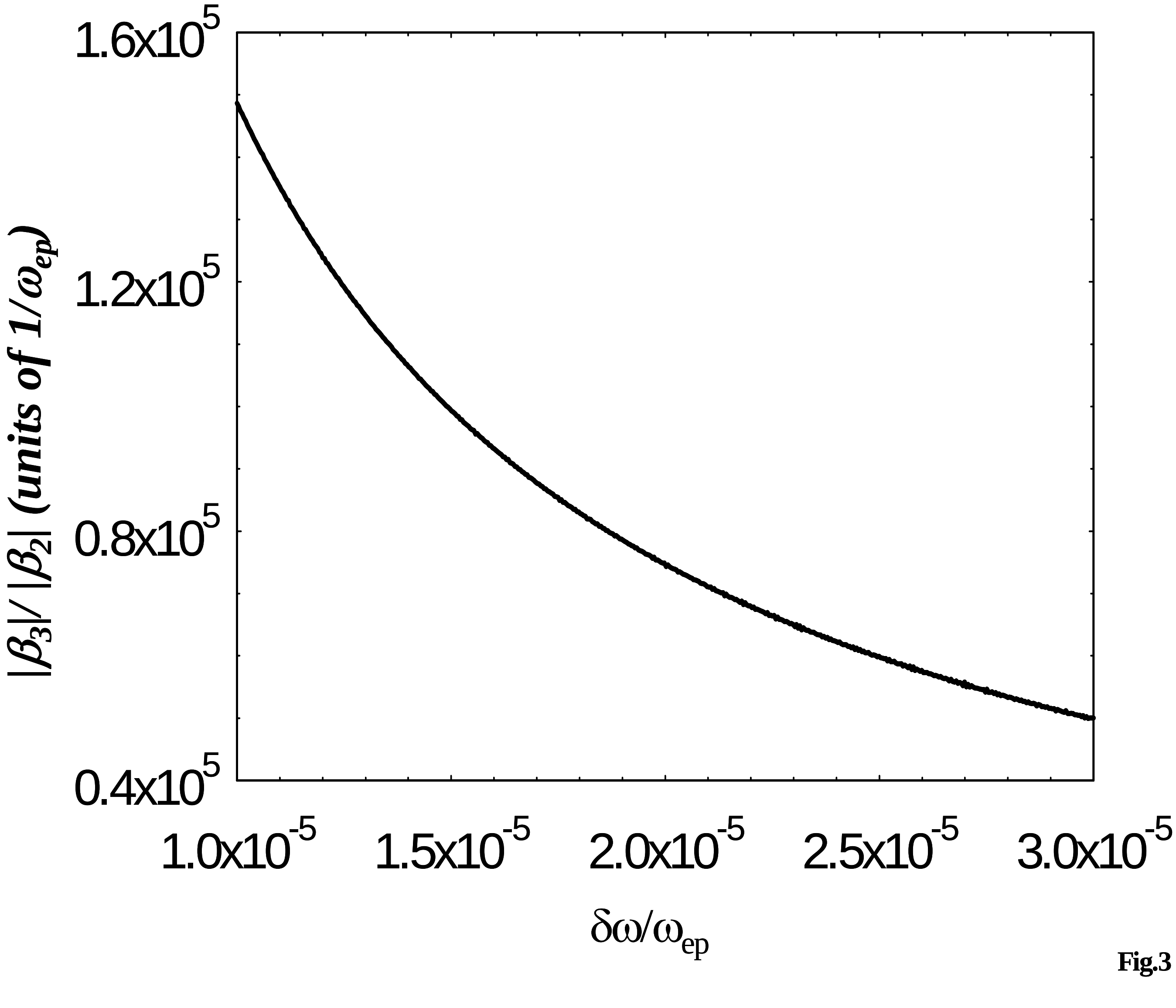

**Fig.3**

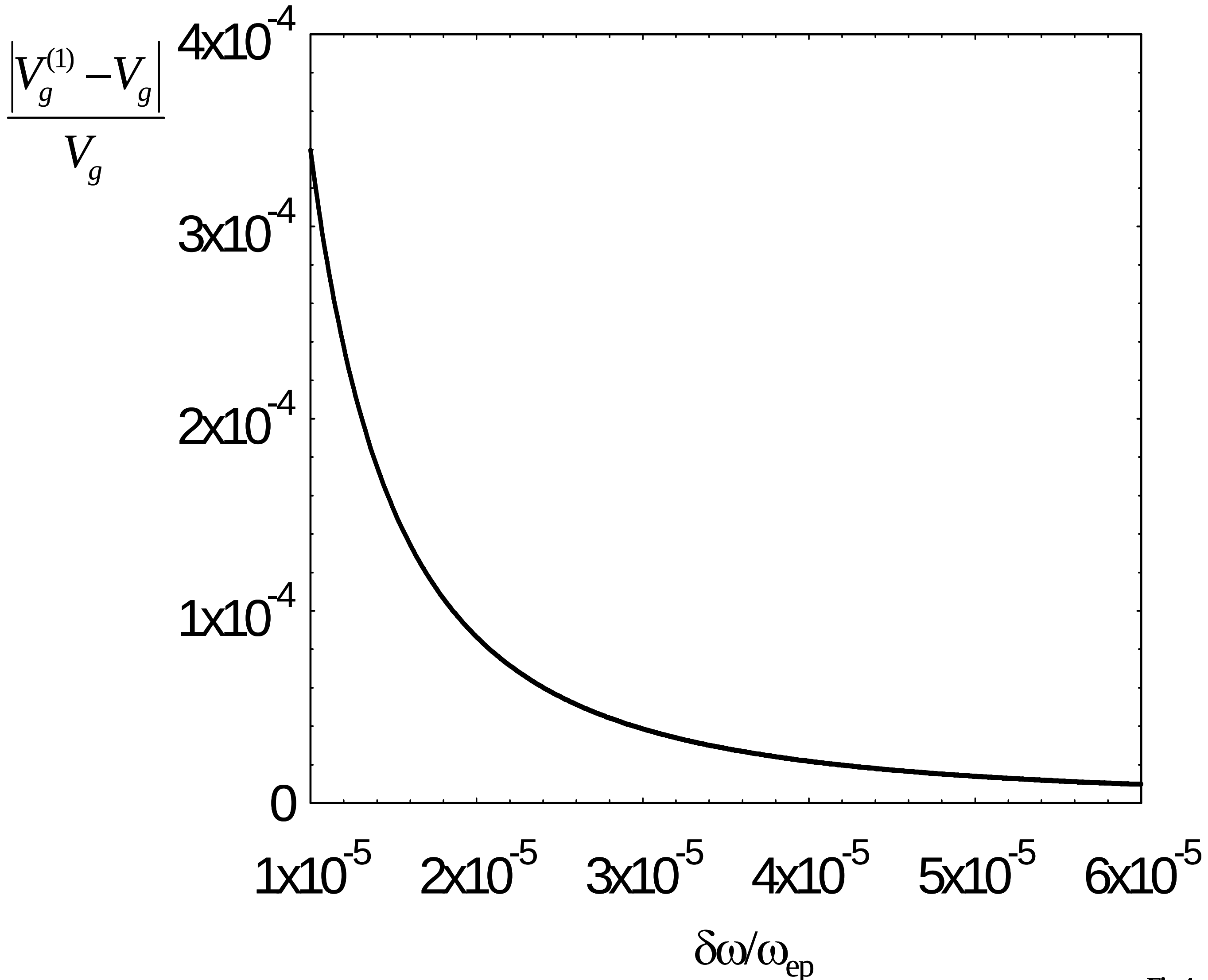

**Fig.4**